\documentclass[a4paper,11pt]{article}
\usepackage{jinstpub} 
\usepackage{lineno}
\usepackage{subfigure}
\usepackage{siunitx}
\usepackage{float}

\title{\boldmath Three-dimensional position reconstruction of orthogonal-strip planar high-purity germanium detectors using maximum likelihood estimation}

\author{Qiuli Zhang\textsuperscript{a,b}, Peng Zhang\textsuperscript{a,b}, Wenhan Dai\textsuperscript{a,b}, Mingxin Yang\textsuperscript{a,b}, Yang Tian\textsuperscript{a,b}, Ming Zeng\textsuperscript{a,b}, Hao Ma\textsuperscript{a,b}, Zhi Zeng\textsuperscript{a,b*}}
\affiliation{\textsuperscript{a}Department of Engineering Physics, Tsinghua University,\\
Beijing 100084, China}
\affiliation{\textsuperscript{b}Key Laboratory of Particle and Radiation Imaging, Ministry of Education,\\
Beijing 100084, China}

\emailAdd{zengzhi@tsinghua.edu.cn}

\abstract{Orthogonal-strip planar high-purity germanium (HPGe) detectors can reconstruct three-dimensional (3D) positions of photon interactions through analysis of parameters extracted from multiple charge signals. The conventional method independently reconstructs positions in each dimension using amplitude-based parameters, leading to noise sensitivity and systematic biases. In this study, we propose a multi-parameter-joint reconstruction method based on maximum likelihood estimation (MLE) which establishes a mapping between pulse shape parameters and corresponding 3D positions. To mitigate the effects of electronic noise, we employ integral-based parameters. The reconstruction performance was evaluated using pulse shape simulations. For 100 keV photons under 1 keV root-mean-square (RMS) electronic noise, the maximum Z reconstruction bias was reduced from 0.4 mm to 0.02 mm in the central region and from 2 mm to 0.15 mm near the electrodes. The maximum reconstruction bias in the X/Y directions was reduced from 0.4 mm to 0.016 mm. Furthermore, the use of integral-based parameters mitigated the rapid degradation of resolution under high-noise conditions. The achieved position resolution ranged from 0.07 mm to 0.16 mm in the Z directions and  from 0.07 mm to 0.44 mm in the X/Y direction. This method offers a promising approach to 3D position reconstruction with HPGe detectors for applications such as medical imaging and gamma-ray astronomy.}


\begin{document}
\maketitle
\flushbottom

\section{Introduction}
\label{sec:intro}

Due to excellent energy resolution and position sensitivity, orthogonal-strip planar high-purity germanium (HPGe) detectors have demonstrated significant application potential in fields such as medical imaging\cite{charac-pet,charac-spect}, gamma-ray astronomy\cite{cosi2020,tomsick2023cosi}, and rare-event searches\cite{0vbb-hpge}. By segmenting the anode into strips and the cathode into orthogonal strips\cite{amorphGe-amman,amorphGe-luke},  such detectors are capable of reconstructing the three-dimensional (3D) interaction positions of photons. Position reconstruction is generally performed by analyzing the pulse shape parameters of multiple charge signals\cite{despec2023,cali-cosi}. The selection of pulse shape parameters and the reconstruction algorithm are crucial for accurate positioning.
\par{}
Conventional method reconstructs each spatial dimension independently. Z position is typically determined by computing the collection time difference (CTD) between signals from the triggered cathode and anode strips\cite{depth-hpge,3dpos-armman}, while X and Y positions are reconstructed based on the amplitude of image signals induced on neighboring strips\cite{positionxy-hpge,positionxy-aganda,pss-yjz}. However, this approach has two major limitations. First, amplitude-based parameters are sensitive to electronic noise, which degrades reconstruction accuracy. Second, independent reconstruction for each dimension ignores cross-dimensional interference, which refers to the phenomenon that reconstructing the position in one dimension is influenced by the event’s positions in the other dimensions. For example, the relationship between CTD and Z position varies with X/Y positions, which will cause systematic biases when we reconstruct the Z position without considering the X/Y positions.
\par{}
In this study, we propose a multi-parameter-joint reconstruction method based on maximum likelihood estimation (MLE). This method aims to achieve unbiased 3D position reconstruction by modeling the relationship between pulse shape parameters and interaction positions, and to mitigate electronic noise effects by optimizing the selection of pulse shape parameters.

\section{Materials and Methods}
\label{sec:method}

\subsection{Detector Structure}

\begin{figure}[H]
    \centering
    \includegraphics[width=0.7\linewidth]{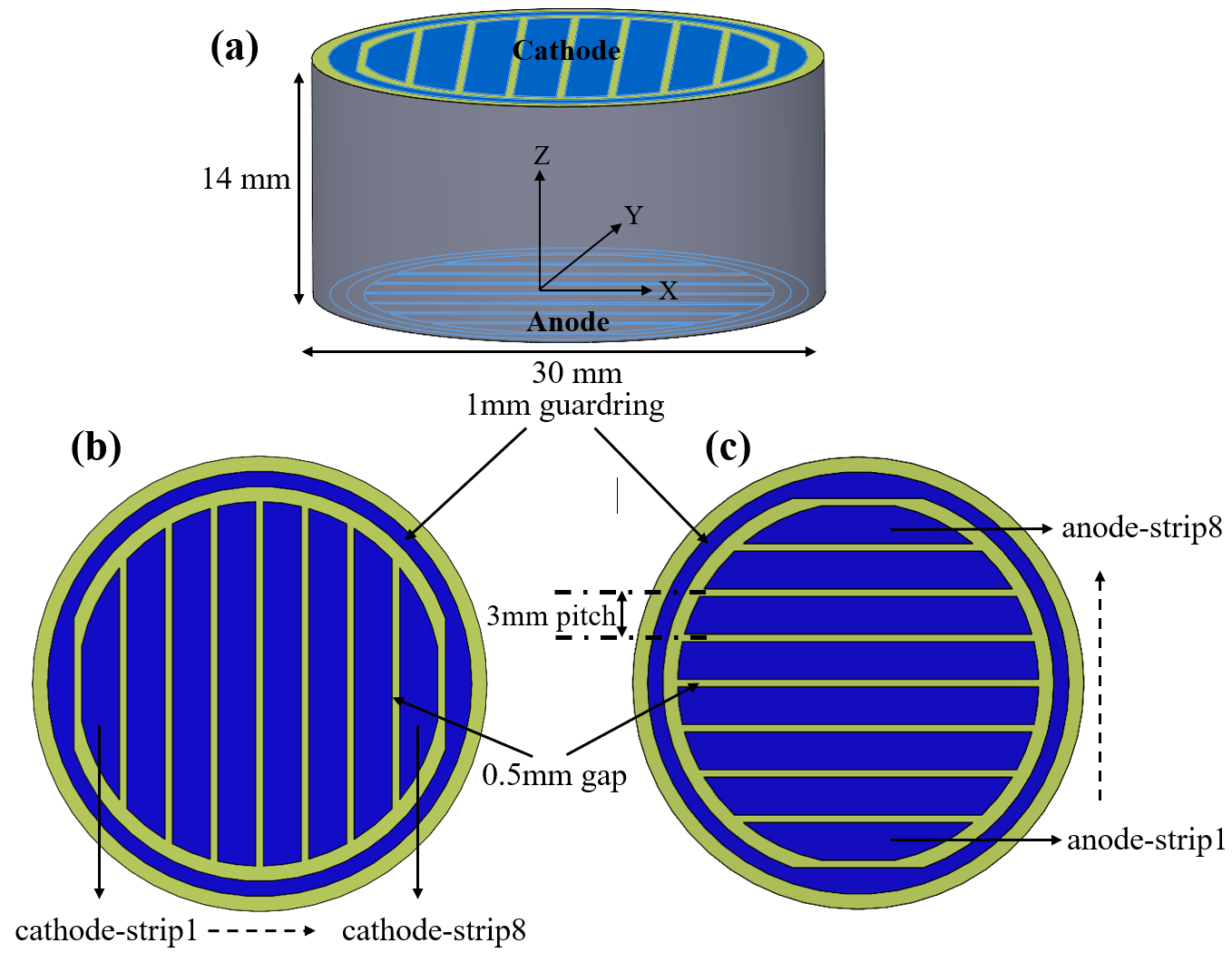}
    \caption{(a) Geometry of the HPGe detector, showing the cylindrical active region (30.0 mm diameter, 14.0 mm thickness); (b) Layout of eight parallel strip electrodes on the cathode (top) plane, all biased to $-1000~\mathrm{V}$, with a 3.0 mm pitch, 0.5 mm gap, and a surrounding 1.0 mm wide guard ring;  (c) Layout of the eight parallel strip electrodes on the anode (bottom) plane, all grounded and oriented orthogonally to the cathode strips (identical geometry parameters).}
    \label{fig:hpge-geo}
\end{figure}

Figure~\ref{fig:hpge-geo} shows the structure of the HPGe detector studied in this work, which features a cylindrical active region with a diameter of 30.0 mm and a thickness of 14.0 mm. Both the top (cathode,$-1000~\mathrm{V}$) and bottom (anode, grounded) planes are patterned with eight parallel strip electrodes. A guard ring with a width of 1.0 mm surrounds the periphery of each plane.

\subsection{Pulse Shape Simulation}
\label{sec:pulse_sim}
Position reconstruction relies on establishing a mapping between pulse shape parameters and their corresponding positions, which requires simulated waveforms with known interaction locations. In this work, the SolidStateDetectors.jl package ~\cite{ssd} is adopted to simulate waveforms

\par{}

\begin{figure}[H]
  \centering
  \includegraphics[height=0.35\hsize]{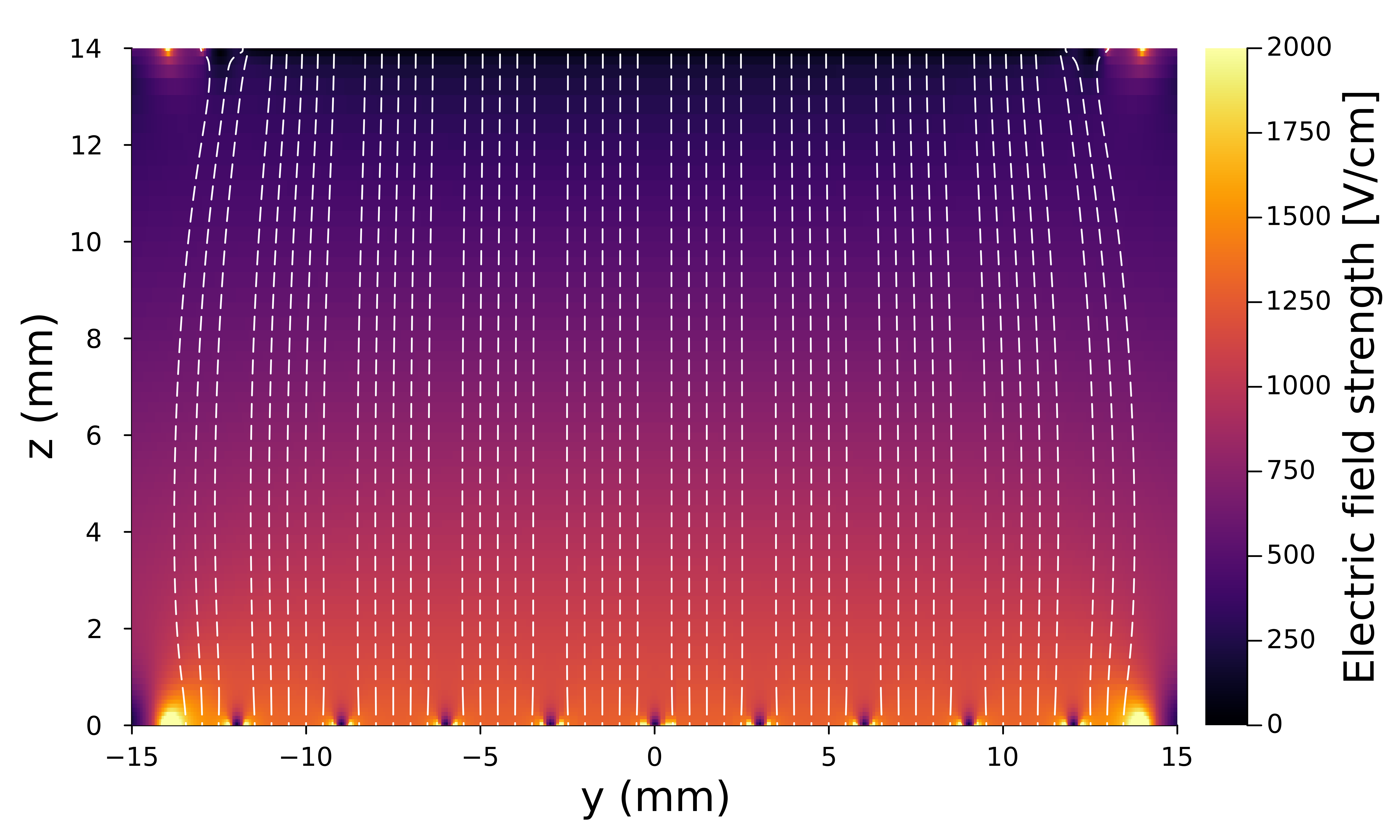}
  \includegraphics[height=0.5\hsize]{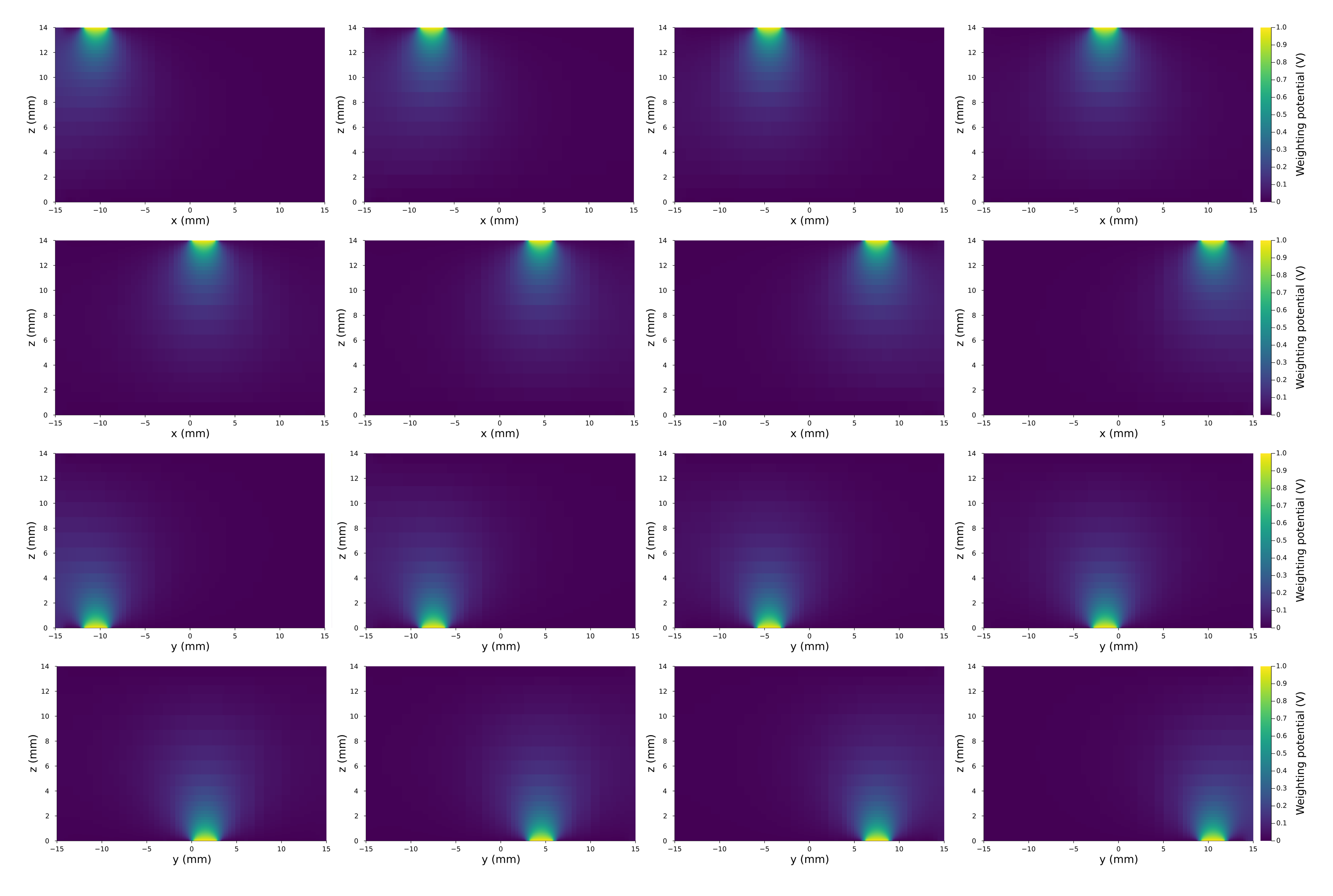}
  \caption{Top: Electric field distribution in the $YZ$ plane, with field lines indicated in white. Bottom: Weighting potential distributions for eight cathode strip electrodes (XZ plane) and eight anode strip electrodes (YZ plane).}
  \label{fig:field}
\end{figure}

Pulse shape simulation consists of two main steps. First, the electric field and weighting potential are calculated using the detector geometry (Figure~\ref{fig:hpge-geo}). The electric field determines the drift trajectories of charge carriers, while the weighting potential is used to compute induced charge signals. The impurity concentration is set to $7.0 \times 10^{9}~\mathrm{cm^{-3}}$  at the detector bottom, with a linear gradient of -2.86\% along the Z direction. The calculated electric field and weighting potential distributions are shown in figure~\ref{fig:field}.
\par{}

\begin{figure}[H]
  \centering
  \includegraphics[height=0.4\hsize]{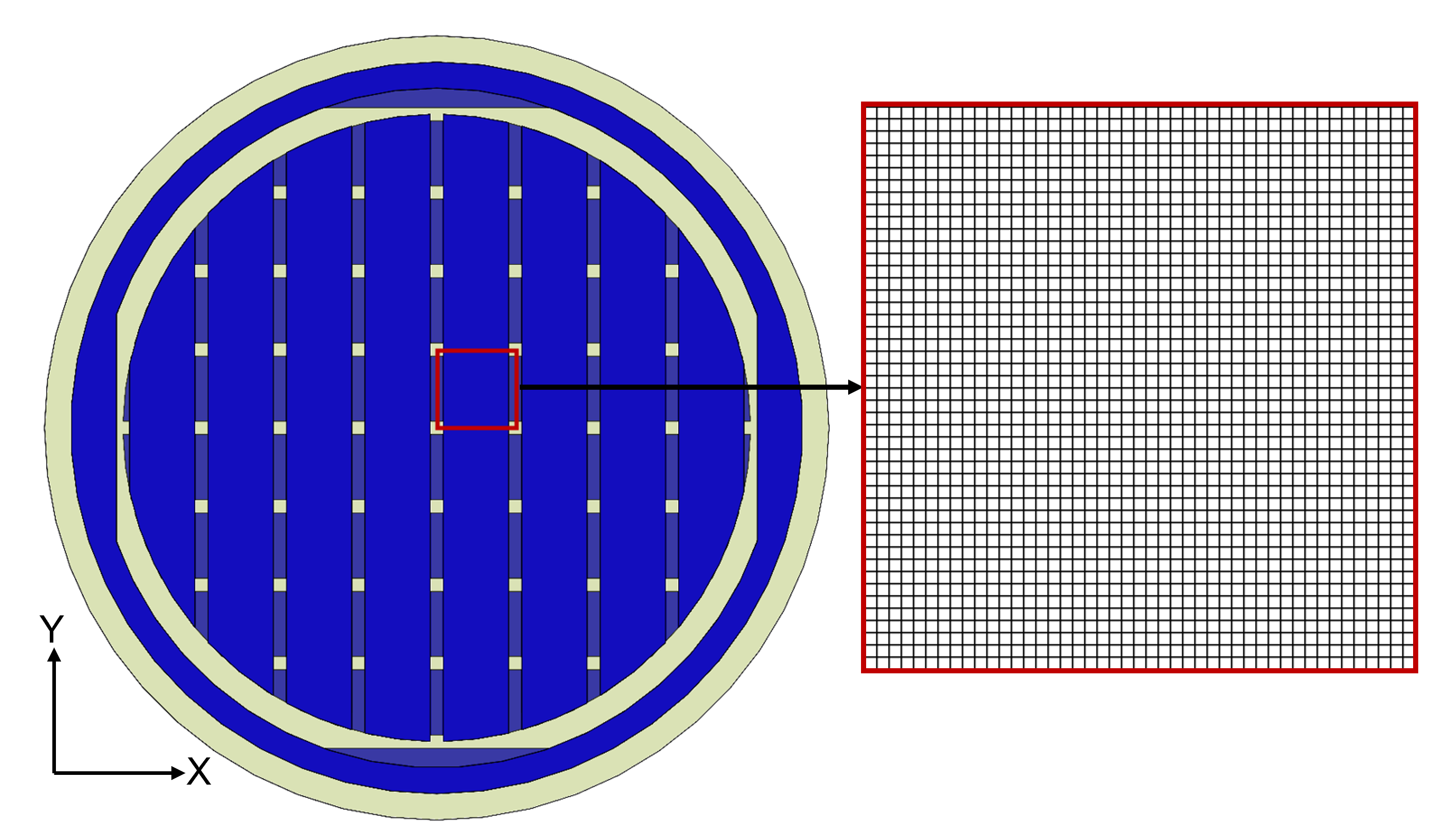}
  \caption{
    Left: Top view of the detector structure, where multiple pixel elements are formed by the intersections of orthogonal anode and cathode strips. The central pixel under study is highlighted by the red box. 
    Right: Magnified view of the selected central pixel, showing the evenly spaced three-dimensional points used to simulate interaction events within this pixel.
  }
  \label{fig:pixel}
\end{figure}

The second step involves simulating the induced charge signals at the detector electrodes, which requires specifying the positions and energies of interaction events. Given the double-sided orthogonal strip electrode structure, the detector is segmented into multiple pixel elements, each defined by the intersection of one anode strip and one cathode strip. Due to the periodicity and symmetry of electrode structure, the induced signal characteristics are essentially identical for most central pixels. In this work, a typical central pixel is simulated to represent all central pixels, while edge pixels are not considered, as illustrated in Figure~\ref{fig:pixel}. Within this central pixel, a set of evenly spaced three-dimensional sampling points with 0.1~mm spacing is generated, with each point corresponding to a simulated interaction event. In the X and Y directions, the sampling region extends from the center of one inter-strip gap to the next, fully enclosing one pixel, while along the Z direction, it spans the entire detector depth. All interaction events are assigned an energy of 100 keV.

\vspace*{\fill}
\clearpage

\par{}

The induced charge signal  $Q_i(t)$  can be determined using the Shockley-Ramo theorem\cite{Shockley,hezhong-Shockley–Ramo,ramo}:

\begin{equation}
\label{eq:shockley_ramo}
Q_i(t) = Q_0 \left[ \phi_i^w (r_h(t)) - \phi_i^w (r_e(t)) \right]
\end{equation}

where $Q_0$ is the absolute electric charge of the charge clouds, and $\phi_i^w(r_{e/h}(t))$ represents the weighting potential of electrode i at position $r_{e/h}(t)$. The effects of diffusion and self-repulsion of charge clouds are simulated for each interaction event. In SolidStateDetectors.jl, diffusion is realized by simulating a random walk for each charge carrier per time step, while self-repulsion is simulated by considering the electric field generated by charge carriers. Simulated pulse shapes are convolved with an RC preamplifier response ($\tau = \SI{14}{\micro\second}$) and gaussian noise.
\par{}

\begin{figure}[H]
  \centering
  \includegraphics[height=0.5\textwidth]{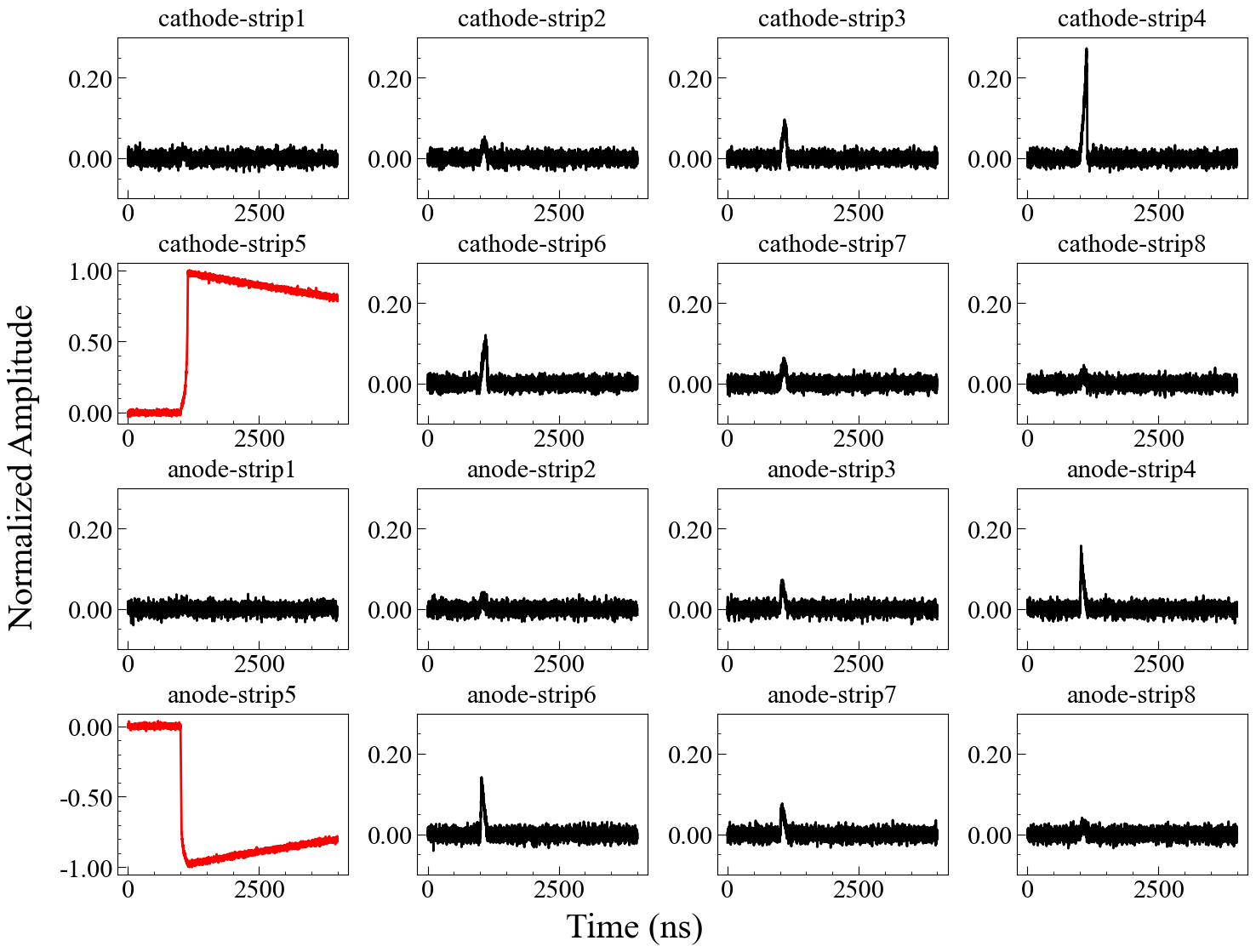}%
  \caption{Simulated normalized pulse shapes from all strip electrodes on the cathode and anode planes for an interaction event. Charge collection signals are shown in red, and image signals are shown in black.}
  \label{fig:example-pulse}
\end{figure}

An example of simulated pulse shapes for an interaction event is shown in figure~\ref{fig:example-pulse}. When a photon interacts within the detector,charge collection signals are induced on the strip electrodes that collect drifting charge carriers, while charge image signals are induced on the remaining strips electrodes. Both types of signals carry spatial information. The amplitude of an image signal decreases with increasing distance from the charge collection electrode. In this study, both the charge collection signals and the neighboring image signals are jointly used to maximize reconstruction accuracy.

\vspace*{\fill}
\clearpage

\subsection{Position-Sensitive Pulse Parameters}

\begin{figure}[H]
  \centering
  \includegraphics[height=0.5\textwidth]{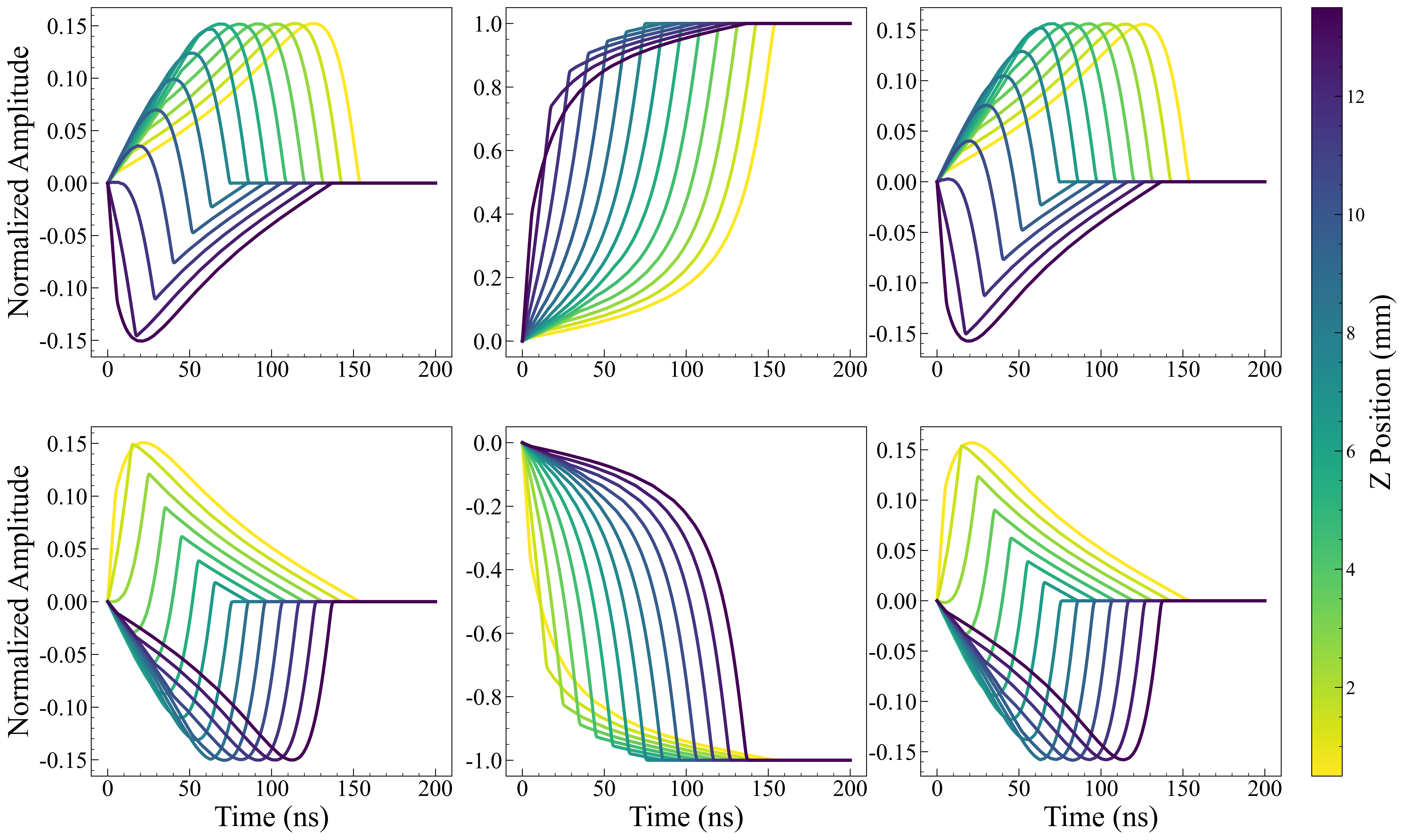}%
  \caption{Simulated normalized pulse shapes at various Z positions with fixed X and Y. The top and bottom rows show signals from the cathode and anode planes, respectively. In each row, the three panels (left to right) display the left neighboring image signal, the charge collection signal, and the right neighboring image signal. The polarity of the image signals and the concavity of the charge collection signal change with Z position.}
  \label{fig:pulse-diffz}
\end{figure}

Figure~\ref{fig:pulse-diffz} shows simulated normalized pulse shapes of charge collection and neighboring electrodes. Here, the X and Y positions of the interaction are fixed, while the Z position varies from 0 mm (anode) to 14 mm (cathode). As the Z position increases, the charge collection pulse shapes change from concave to convex, and the image signals undergo a polarity transition from positive, through bipolar, to negative. These Z-dependent features are essential for Z position reconstruction via pulse shape analysis. To exploit these features and suppress electronic noise, we define two integral-based parameters for Z position reconstruction using the algebraic sum of the net areas.
\par{}
The first parameter, $S_C$, is defined based on the charge collection signals:

\begin{equation}
\label{eq:sc}
S_C = S_{\text{AN}} + S_{\text{CA}}
\end{equation}
where $S_{\text{AN}}$ and $S_{\text{CA}}$ represent the net areas of the anode and cathode charge collection signals, respectively.
\par
\begin{figure}[H]
  \centering
  \includegraphics[height=0.6\textwidth]{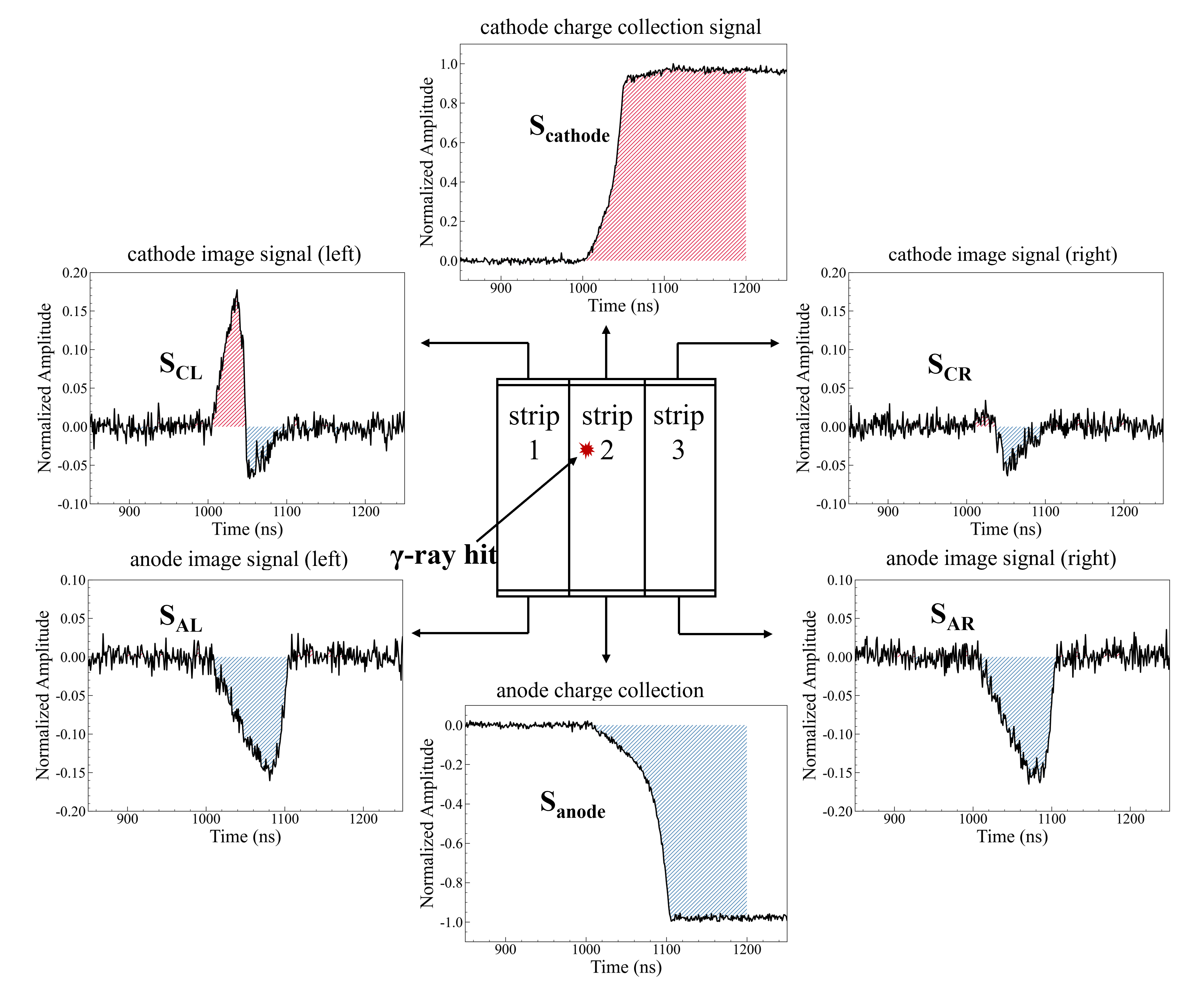}%
  \caption{ Schematic diagram of net area calculation for an interaction event. For Z position reconstruction, the net areas of charge collection signals and neighboring image signals from the anode and cathode planes are calculated. The integration window for each signal is 300 ns, from 900 ns to 1200 ns. Areas above the baseline are integrated as positive (red), those below as negative (blue).}
  \label{fig:net-area}
\end{figure}

\begin{figure}[H]
\centering
\includegraphics[height=0.33\textwidth]{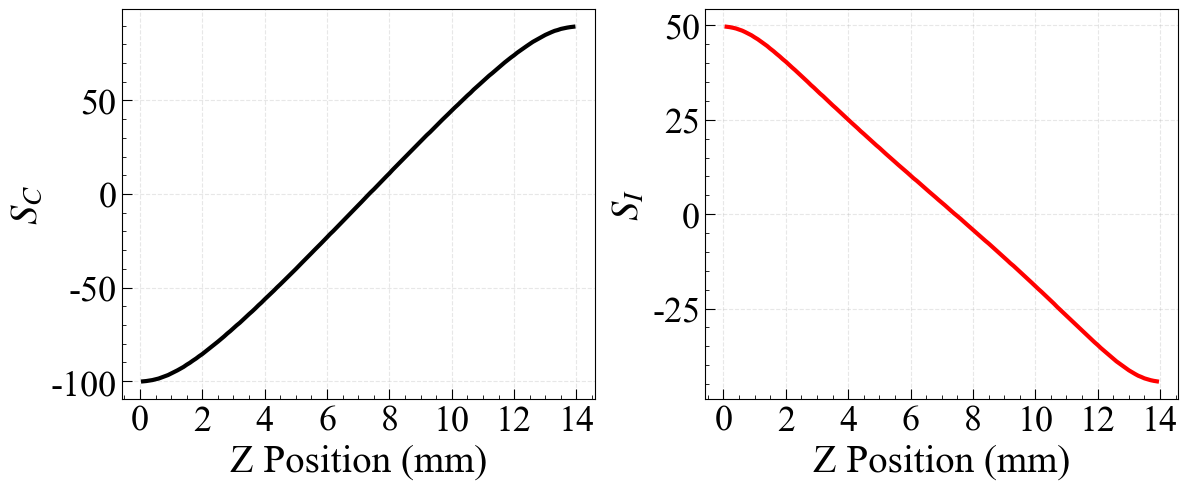}
\caption{ $S_{\text{C}}$ (left) and $S_{\text{I}}$  (right) as functions of Z position. Both parameters exhibit strong linearity in the central region of the detector, with deviations observed near the electrode surfaces.}
\label{fig:SC-SI-z}
\end{figure}

{}
The second parameter, $S_I$, is defined based on the image signals induced on the neighboring electrodes:

\begin{equation}\label{eq:si}
S_I = S_{\text{AN,L}} + S_{\text{AN,R}} + S_{\text{CA,L}} + S_{\text{CA,R}}
\end{equation}
where $S_{\text{AN,L}}$, $S_{\text{AN,R}}$, $S_{\text{CA,L}}$ and $S_{\text{CA,R}}$ denote the net areas of the left/right image signals on the anode/cathode planes, respectively, as illustrated in Figure~\ref{fig:net-area}. The net area for each signal is obtained by integrating the waveform over a 300 ns window which covers the whole rise edge. Areas above the baseline are treated as positive, while areas below the baseline are treated as negative. The integration window should exceed the maximum carrier collection time to ensure complete charge integration. As shown in Figure~\ref{fig:SC-SI-z}, both $S_{\text{C}}$ and $S_{\text{I}}$ exhibit linear relationships with Z position except near the electrodes, making Z position reconstruction possible.

\par{}

\begin{figure}[H]
  \centering
  \includegraphics[height=0.28\textwidth]{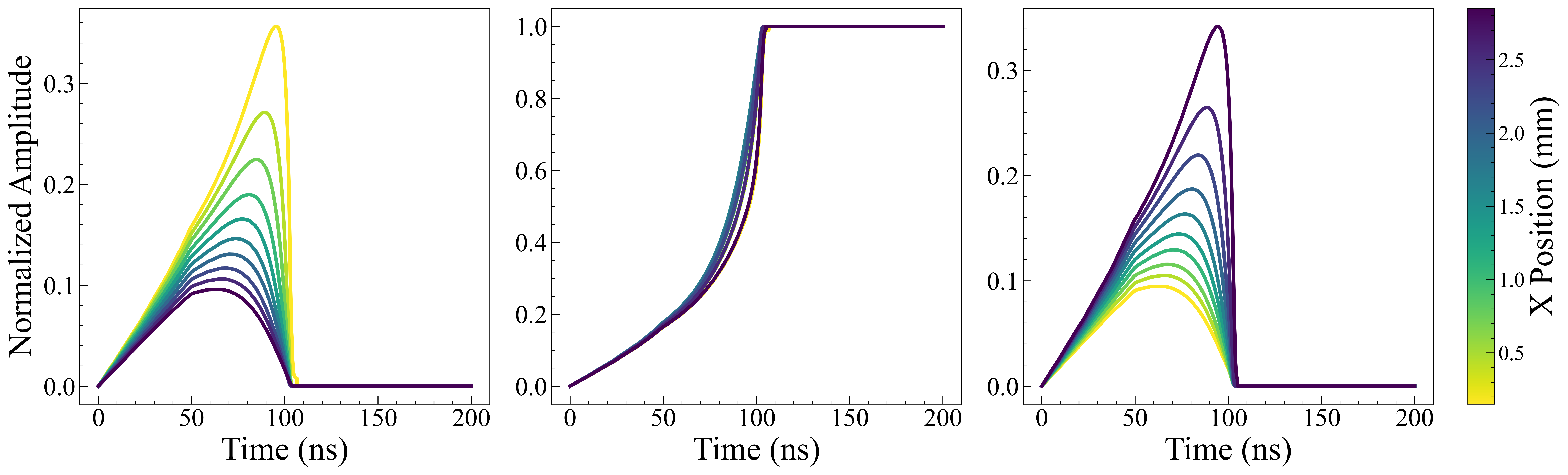}%
  \caption{Simulated normalized pulse shapes at various X positions with fixed Y and Z. The three panels (left to right) display the left neighboring image signal, the charge collection signal, and the right neighboring image signal. The amplitudes of the left and right image signals vary inversely with the X position.}
  \label{fig:pulse-diffx}
\end{figure}

\begin{figure}[H]
  \centering
  \includegraphics[height=0.41\textwidth]{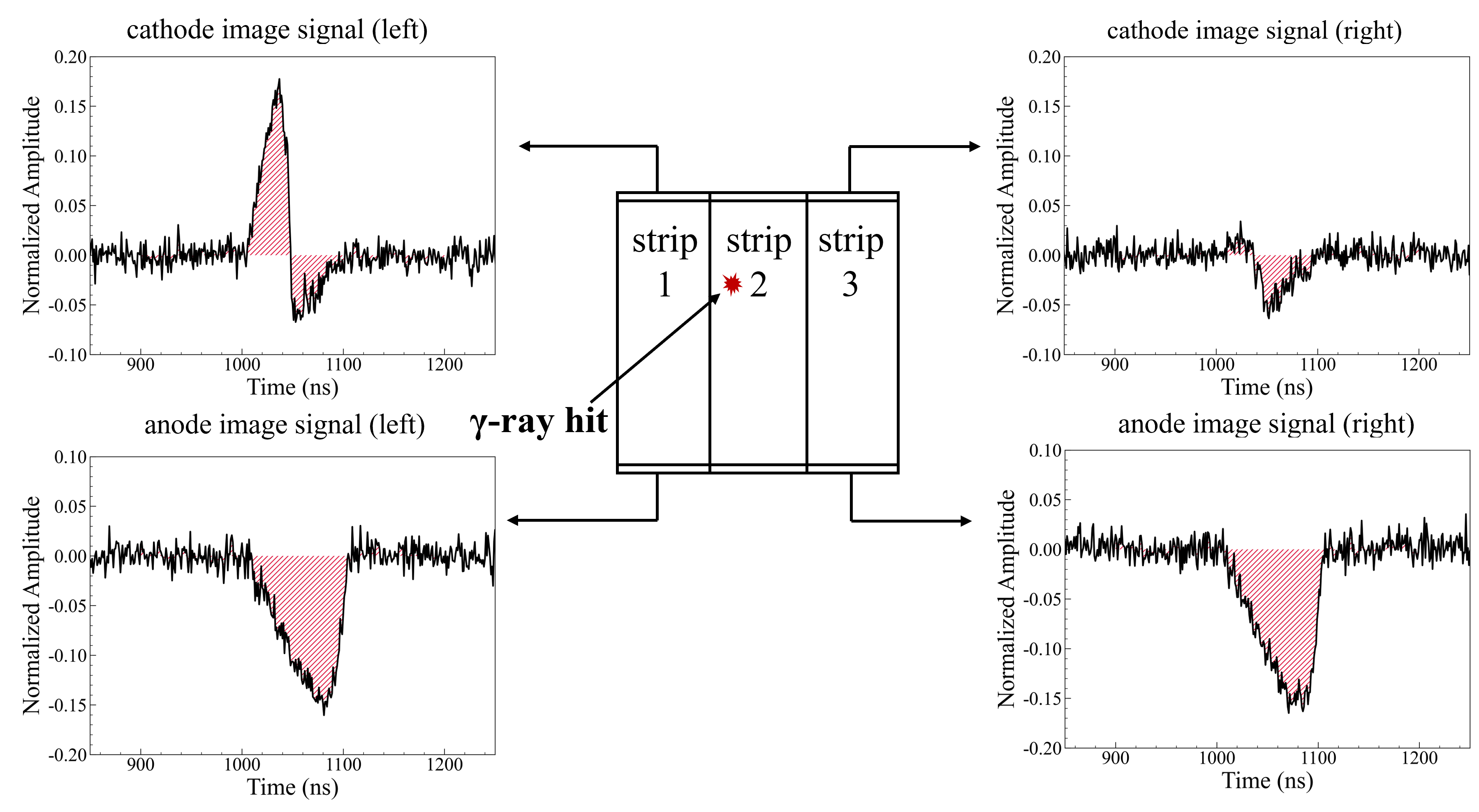}%
  \caption{Schematic diagram of absolute area calculation for an interaction event. For X and Y position reconstruction, the absolute areas of the left and right neighboring image signals from the anode and cathode planes are calculated. The absolute value of the waveform is integrated within a 300 ns window from 900 ns to 1200 ns, and the integration region is shown in red.}
  \label{fig:asb-area}
\end{figure}

The simulated normalized pulse shapes of the neighboring electrodes at different X positions (with Y and Z positions fixed)  are shown in Figure~\ref{fig:pulse-diffx}. As the interaction position moves laterally across the strip, the amplitude of the neighboring image signal closer to the interaction increases, while that of the farther one decreases. This asymmetry characteristic offers position-sensitive information for spatial reconstruction. The conventional method uses the parameter $A_{\text{asy}}$ for X and Y position reconstruction, defined as

\begin{equation}\label{eq:aasy}
A_{\text{asy}} = \frac{A_{\text{L}} - A_{\text{R}}}{A_{\text{L}} + A_{\text{R}}}
\end{equation}

where $A_{\text{L}}$ and $A_{\text{R}}$ are the amplitudes ( the difference between the maximum and minimum values) of the image signals from the left and right neighboring electrodes, respectively. However, the amplitude-based parameter is sensitive to electronic noise. To address this problem, we introduce an integral-based parameter $S_{A}$ for the X and Y positions reconstruction, defined as

\begin{equation}\label{eq:sa}
S_{A} = \frac{S_{\text{L}} - S_{\text{R}}}{S_{\text{L}} + S_{\text{R}}}
\end{equation}
where  $S_{\text{L}}$ and $S_{\text{R}}$ are the absolute areas of the neighboring image signals. The absolute area is calculated by integrating the absolute value of the charge signal over the same 300 ns time window shown in figure~\ref{fig:asb-area}. 
\par
\begin{figure}[H] 
\centering 
\includegraphics[height=0.32\textwidth]{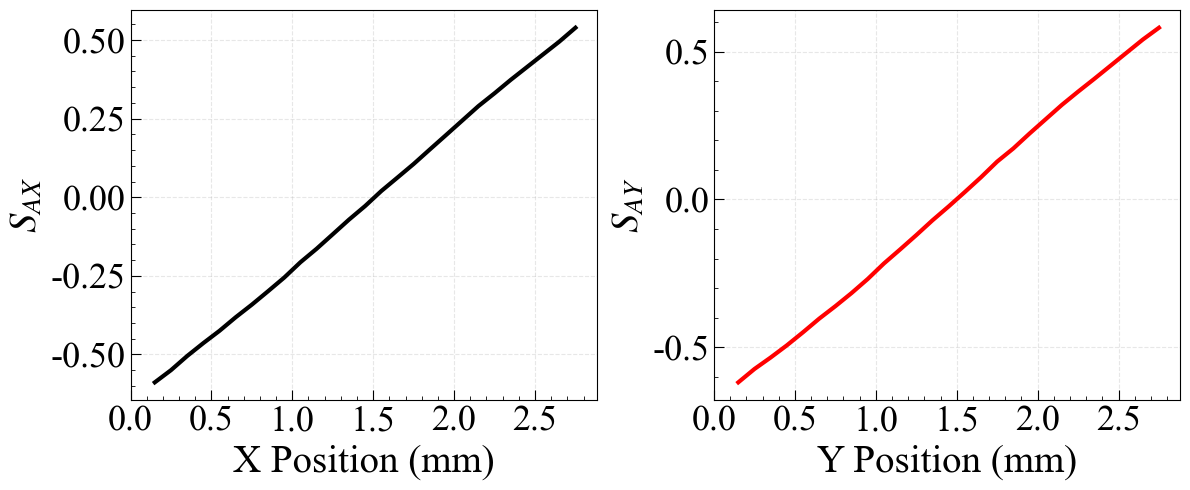} 
\caption{ $S_{\text{AX}}$ (left) as a function of the X postion and $S_{\text{AY}}$(right) as a function of the Y postion, both showing good linear relationships across the strip.} 
\label{fig:SAX-SAY-xy} 
\end{figure}

{}
Given the double-sided orthogonal strip electrode design, the parameter $S_{\text{AX}}$, calculated from the image signals on the cathode plane, is used for X position reconstruction, while $S_{\text{AY}}$, obtained from the image signals on the anode plane, is used for Y position reconstruction. Both $S_{\text{AX}}$ and $S_{\text{AY}}$ exhibit a clear linear dependence on the corresponding spatial coordinates, as demonstrated in Figure~\ref{fig:SAX-SAY-xy}.

\vspace*{\fill}
\clearpage

\subsection{Maximum Likelihood Estimation for 3D Position Reconstruction}

To address cross-dimensional interference in position reconstruction, we propose a multi-parameter-joint reconstruction method based on MLE. A set of integral-based parameters, $\mathcal{A} = \{S_C, S_I, S_{AX}, S_{AY}\}$ is employed. We establish a mapping between the integral-based parameters and the corresponding 3D positions. Specifically, for each interaction event described in Section~\ref{sec:pulse_sim}, independent noise  are added to the raw pulses to generate $\mathcal{N}$ sets of parameters $\mathcal{A}$. These parameters are modeled using a multivariate normal distribution. The sample mean and covariance matrix are estimated as follows:

\begin{equation}\label{eq:mu}
\hat{\boldsymbol{\mu}} = \frac{1}{N}\sum_{i=1}^N \mathcal{A}_i
\end{equation}

\begin{equation}\label{eq:sigma}
\hat{\boldsymbol{\Sigma}} = \frac{1}{N}\sum_{i=1}^N (\mathcal{A}_i - \hat{\boldsymbol{\mu}})(\mathcal{A}_i - \hat{\boldsymbol{\mu}})^T
\end{equation}

The estimated position corresponds to the coordinates that maximize the likelihood function: 
\begin{equation}\label{eq:mle}
(\hat{x}, \hat{y}, \hat{z}) = \mathop{\arg\max}\limits_{(x,y,z)} \mathcal{P}\left(\mathcal{A}\ |\ \boldsymbol{\mu}(x,y,z), \boldsymbol{\Sigma}(x,y,z)\right)
\end{equation}
Here, $\mathcal{P}$  denotes the probability density function of a multivariate normal distribution with mean $\mathbf{\mu}$ and covariance matrix $\mathbf{\Sigma}$. 
\par{}

To evaluate reconstruction performance, multiple independent reconstructions are carried out. The reconstruction bias is quantified by the difference between the mean of estimated positions and the true position. The position resolution along each coordinate axis ($x$, $y$, and $z$) is quantified by the standard deviation of the estimated positions:
\begin{equation}\label{eq:position_resolution_x}
\sigma_X = \sqrt{\frac{1}{M-1} \sum_{j=1}^{N} \left(\hat{x}_j - \bar{\hat{x}}\right)^2}
\end{equation}
where $M$ is the number of repeated reconstructions, $\hat{x}_j$ is the $x$-coordinate estimate of the $j$-th reconstruction, and  $\bar{\hat{x}} = \frac{1}{N} \sum_{j=1}^N \hat{x}_j$ is the sample mean of the $x$-coordinate estimates. Analogous definitions apply to the $Y$ and $Z$ directions to yield $\sigma_Y$ and $\sigma_Z$, respectively.

\section{Results and Discussion}
\subsection{Position Reconstruction Bias}

\begin{figure}[H]
  \centering
  \includegraphics[height=0.65\textwidth]{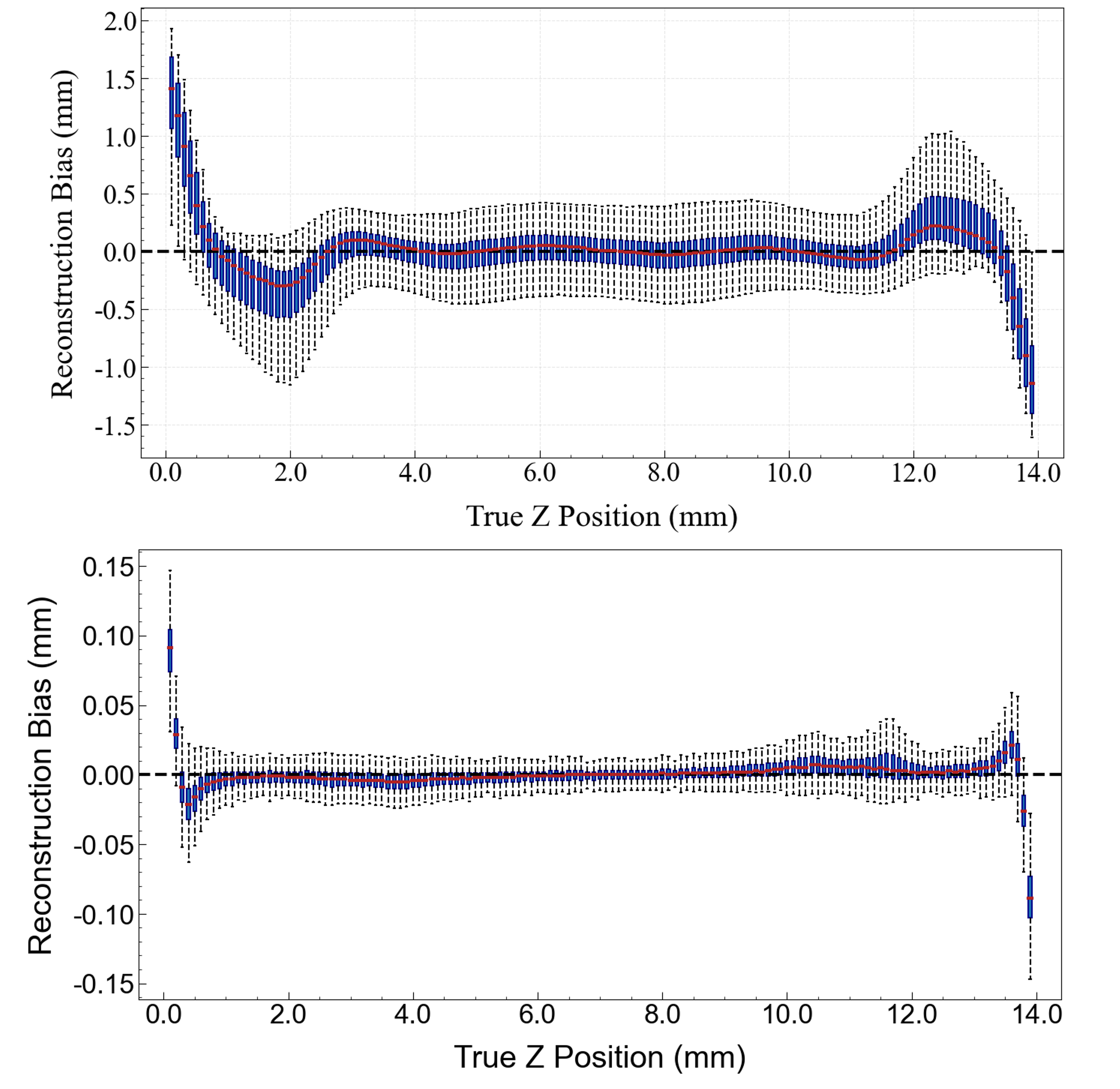}%
  \caption{Boxplots of position reconstruction bias along the Z direction for the conventional method (top) and the MLE method (bottom). At each Z position, all interaction events in the corresponding X-Y plane are used to calculate the bias distribution. For each boxplot: the horizontal red line indicates the median; the box represents the interquartile range (IQR, 25th to 75th percentiles); and the whiskers extend to 1.5 times the IQR from the box edges}
  \label{fig:z-bias}
\end{figure}

Figure ~\ref{fig:z-bias} shows the reconstruction biases along the Z direction for both the conventional and MLE methods. Each boxplot represents the distribution of reconstruction biases at a specific Z position, using results from all X and Y positions. In the central region, both methods exhibit zero-centered symmetric  bias distributions.For the conventional method, the interquartile range (IQR) is approximately ±0.1 mm, which means that $50\%$  of the region at each Z has reconstruction biases within this range, and the maximum bias reaches 0.4 mm. In contrast, the MLE method maintains the IQR within ±0.01 mm, and the maximum bias does not exceed 0.03 mm. 
\par{}
Near the electrodes in the Z direction, both methods show a significant increase in reconstruction bias. This is mainly due to the non-monotonic relationship between the Z position and parameters such as the CTD (used in the conventional method) and the SC and SI (proposed in this work). The polynomial fitting model adopted by the conventional method couldn’t handle the non-linear relationship, resulting in a maximum bias of about 2 mm. By contrast, the maximum bias of the MLE method is about 0.15 mm.
\par
\begin{figure}[H]
  \centering
  \includegraphics[height=0.4\textwidth]{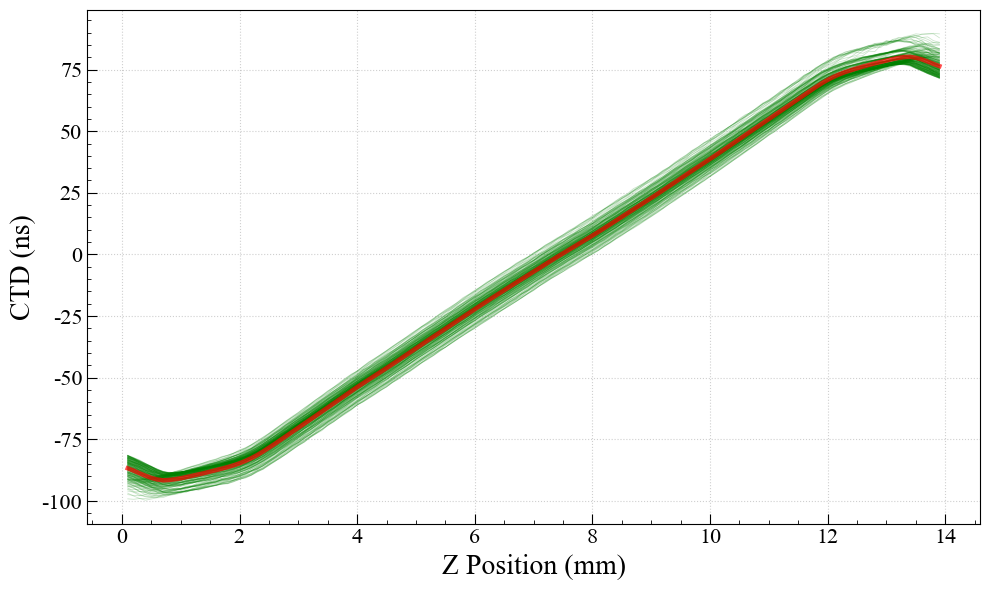}%
  \caption{The dependence of the pulse shape parameter CTD on the Z position. The green lines display individual profiles obtained at various X and Y positions. The red line represents the average CTD over the XY plane for each Z position.}
  \label{fig:CTD-diffxy}
\end{figure}

{}
The conventional method for Z position reconstruction suffers from significant bias due to its simplified modeling of the CTD-Z relationship. As shown in Figure~\ref{fig:CTD-diffxy}, although CTD mainly depends on the Z position, it is also dependent on X and Y positions. The conventional method, which averages CTD values over all X and Y to create a single CTD-Z function, ignores cross-dimensional interference and introduces systematic biases. In contrast, the MLE method maps pulse shape parameters to 3D positions, leading to position reconstruction with much smaller biases.

\vspace*{\fill}
\clearpage

\par
\begin{figure}[H]
  \centering
  \includegraphics[height=0.7\textwidth]{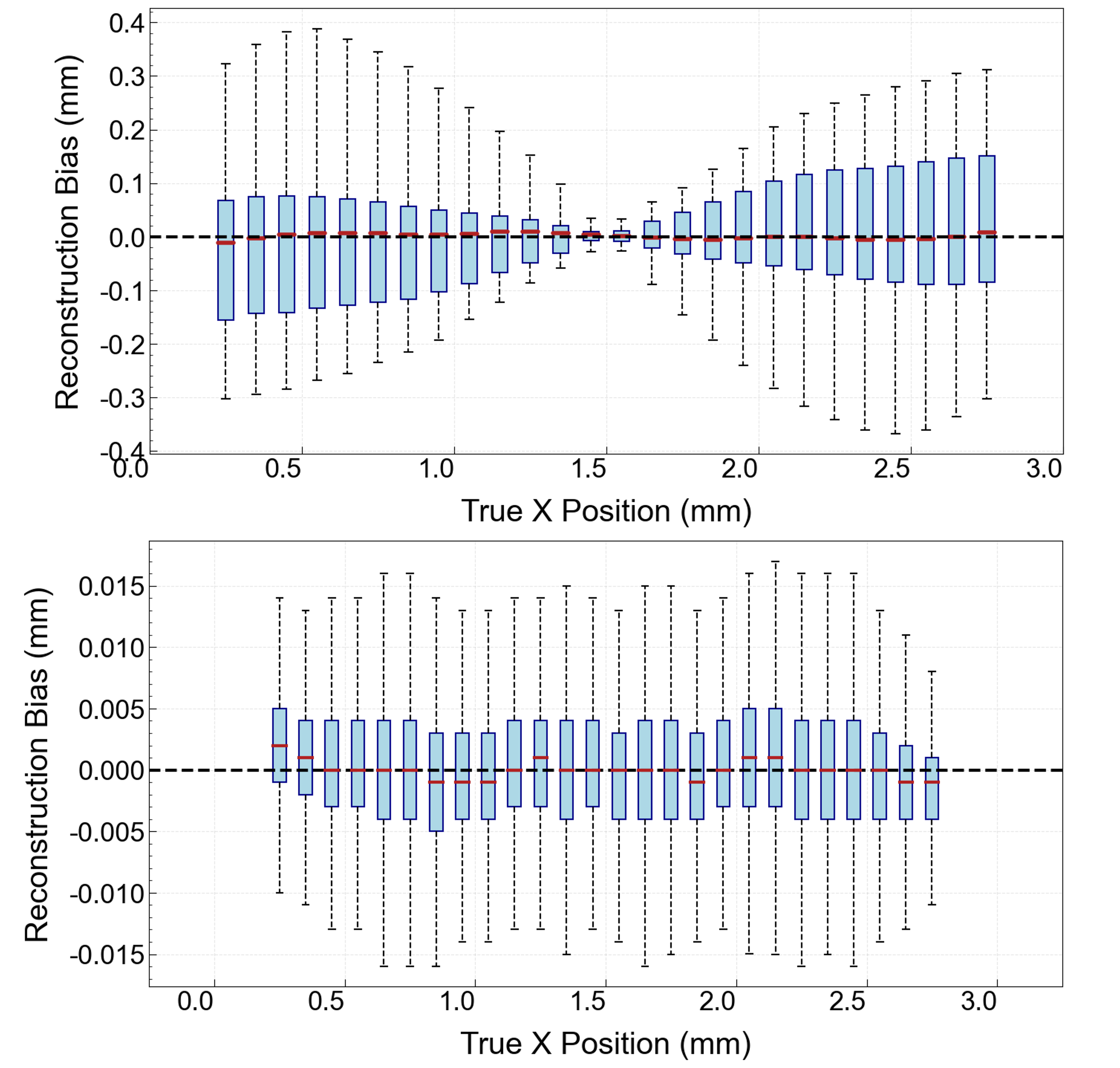}%
  \caption{Boxplots of position reconstruction bias along the X direction for the conventional method (top) and the MLE method (bottom). At each X position, all interaction events in the corresponding Y-Z plane are used to calculate the bias distribution. For each boxplot: the horizontal red line indicates the median; the box represents the interquartile range (IQR, 25th to 75th percentiles); and the whiskers extend to 1.5 times the IQR from the box edges.}
  \label{fig:x-bias}
\end{figure}

\begin{figure}[H]
  \centering
  \includegraphics[height=0.4\textwidth]{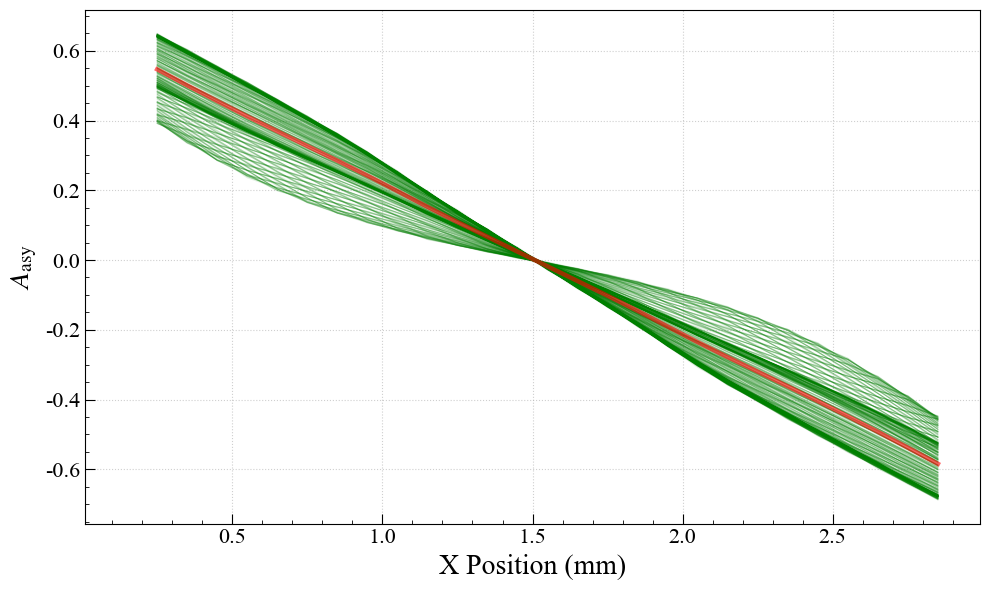}%
  \caption{The dependence of the pulse shape parameter $A_{\text{asy}}$ on the X position. The green lines show individual profiles obtained at various Y and Z positions. The red line represents the average 
$A_{\text{asy}}$ over the YZ plane for each X position.}
  \label{fig:Asym-diffyz}
\end{figure}

{}
Similarly, for X position reconstruction, figure~\ref{fig:x-bias} shows that both methods exhibit zero-centered bias distributions. Using the conventional method, the maximum reconstruction bias increases from 0.05 mm at the strip center to about 0.3 mm near the edges. By contrast, the MLE method maintains the reconstruction bias within 0.015 mm across the entire strip. The larger reconstruction bias in the conventional method arises from cross-dimensional interference affecting the parameter $A_{\text{asy}}$.  Figure~\ref{fig:Asym-diffyz} illustrates this point further, showing that $A_{\text{asy}}$ is not only determined by the X position, but also influenced by the Y and Z positions.

\vspace*{\fill}
\clearpage

\vspace*{\fill}
\clearpage

\subsection{Position Resolution}
\begin{figure}[H]
  \centering
  \subfigure[]{
    \includegraphics[height=0.32\textwidth]{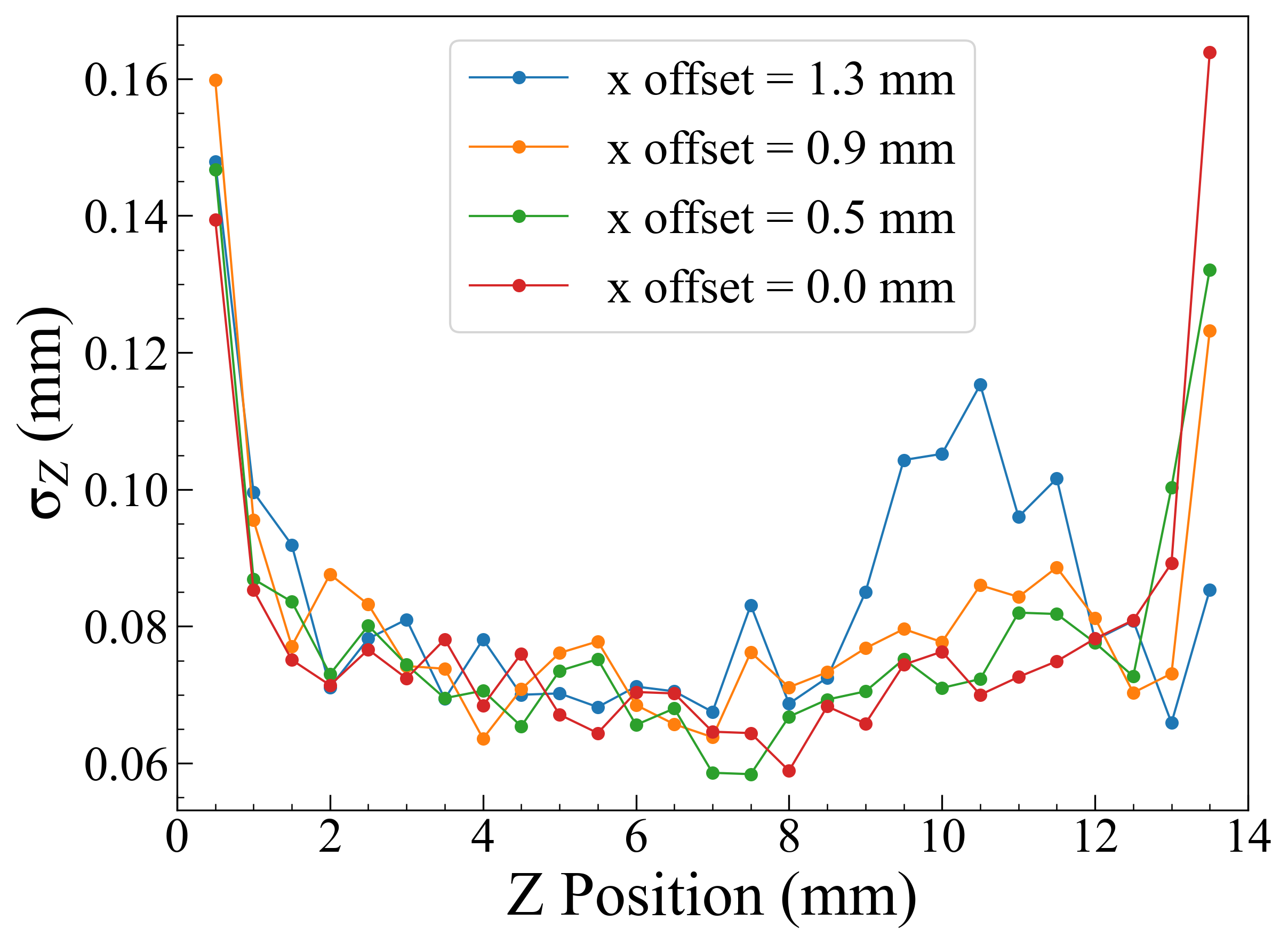}
  }
  \subfigure[]{
    \includegraphics[height=0.32\textwidth]{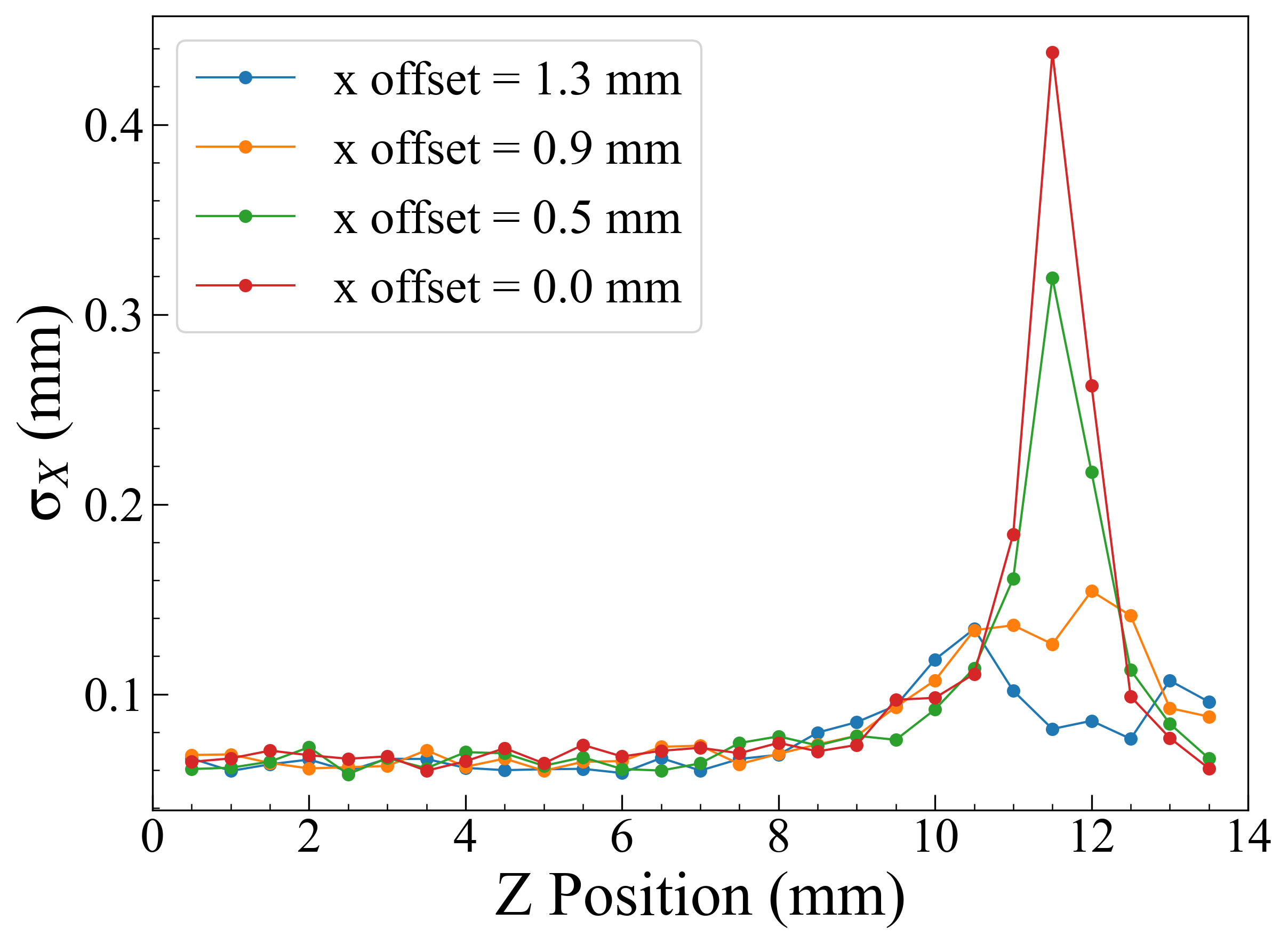}
  }
   \caption{Position resolution performance obtained using the MLE reconstruction method: (a) Z position resolution as a function of Z position, and (b) X position resolution as a function of Z position. The x offset represents the distance from the strip electrode center along the X direction.}
  \label{fig:resolution-mle}
\end{figure}

The MLE reconstruction method demonstrates consistent depth resolution performance across most of the detector volume. As shown in Figure~\ref{fig:resolution-mle} (a), the Z position resolution maintains approximately 0.07 mm throughout the central detector region, regardless of lateral position. However, degradation occurs near both electrode surfaces, where the resolution decreases to approximately 0.16 mm. This degradation results from the strong electric field and large weighting potential gradients near the electrodes, which lead to rapid charge collection and generate signals with steep rising edges. Consequently, the position sensitivity of pulse shape parameters $S_{\text{C}}$ and $S_{\text{I}}$ is reduced, resulting in poorer position resolution near the electrodes.

\par{}
In contrast to the depth resolution, the lateral position resolution exhibits spatial non-uniformity, as demonstrated in Figure~\ref{fig:resolution-mle} (b). The X position resolution maintains stable performance of approximately 0.07 mm in regions near the anode and far from the cathode surface. However, a degradation zone emerges within 1-3.5 mm beneath the cathode electrode, where the resolution deteriorates to approximately 0.44 mm—nearly an order of magnitude worse than other regions. This degradation is attributed to the diminished weighting potential gradient in this region, which renders the induced image signals less distinguishable and reduces the discriminating power of the pulse shape parameter $S_{\text{AX}}$ for lateral position determination.

\begin{figure}[H]
  \centering
  \subfigure[]{
    \includegraphics[height=0.32\textwidth]{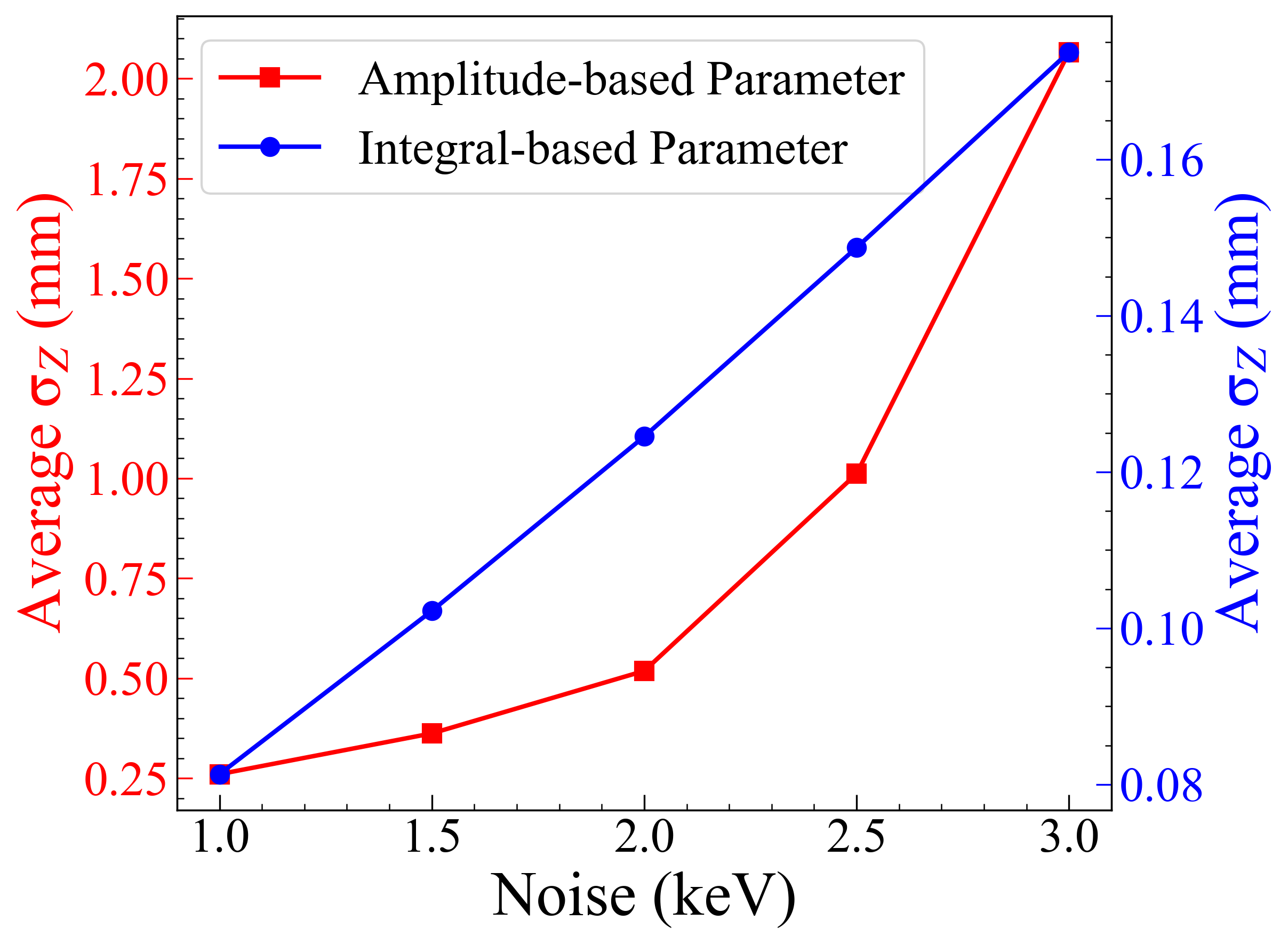}
  }
  \subfigure[]{
    \includegraphics[height=0.32\textwidth]{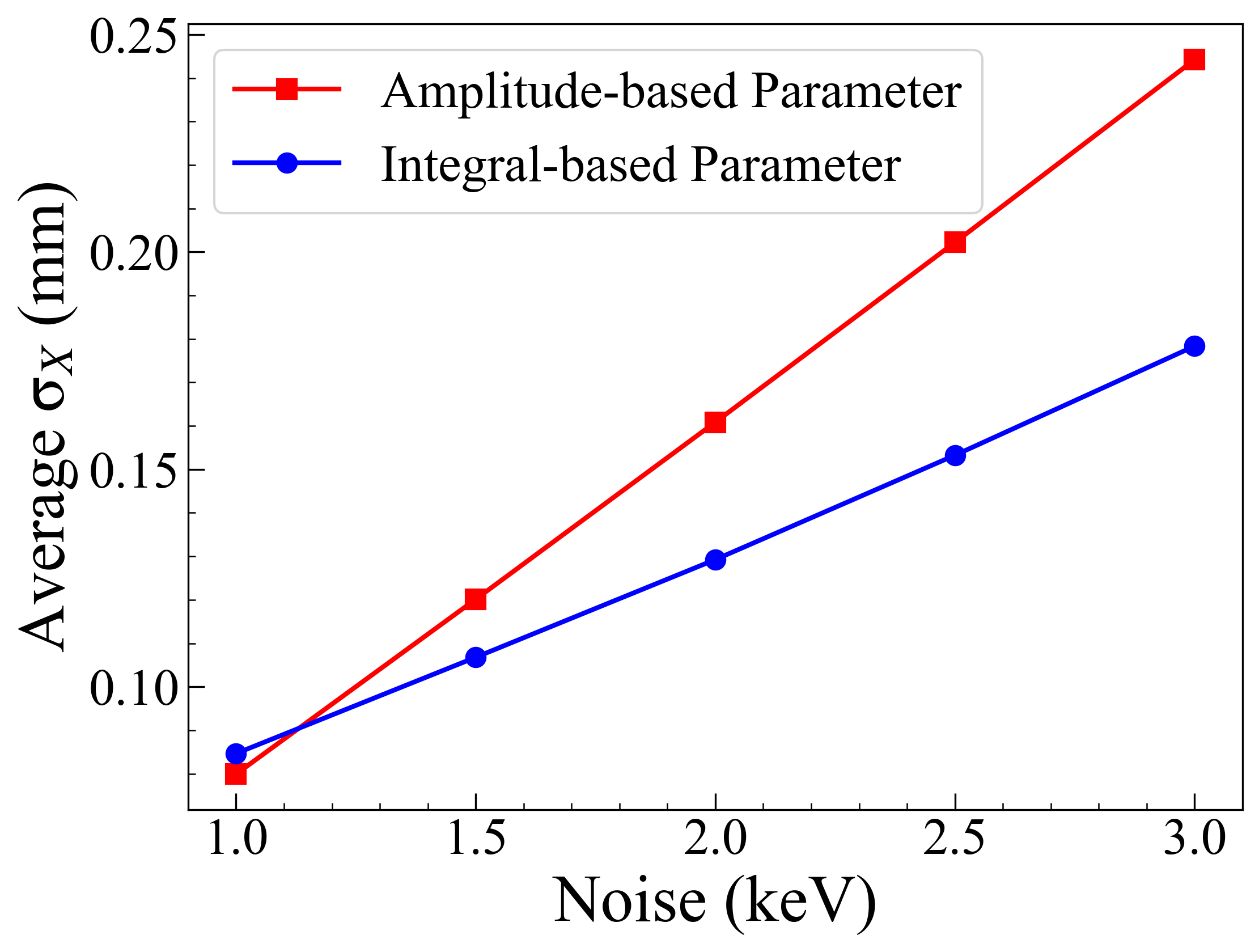}
  }
\caption{(a) Average Z position resolution as a function of noise level using amplitude-based (red) and integral-based (blue) parameters. (b)  Average X position resolution as a function of noise level using amplitude-based (red) and integral-based (blue) parameters. Distinct degradation trends are observed for different parameter types in both panels.}
  \label{fig:resolution-diffnoise}
\end{figure}

For Z position reconstruction, the conventional method uses the amplitude-based parameter (CTD), while the MLE method employs the integral-based parameters ($S_{\text{C}}$ and $S_{\text{I}}$). Figure~\ref{fig:resolution-diffnoise} (a) shows the average Z position resolution at different noise levels, where the average is calculated from the Z position resolution of all interaction events. When using the integral-based parameters, the average Z position resolution degrades linearly, worsening from 0.08 mm to 0.17 mm as the noise RMS increases from 1 keV to 3 keV. In contrast, when using the amplitude-based parameter, the resolution degrades exponentially, increasing rapidly from 0.26 mm to 2.07 mm over the same noise range. This marked difference arises from the large uncertainty in rise time extraction under high noise conditions.

\par{}

Similarly, figure~\ref{fig:resolution-diffnoise} (b) compares the average X position resolution of both the conventional and MLE methods under different noise levels. The conventional method utilizes the amplitude-based parameter ($A_{\text{asy}}$), while the MLE method employs the integral-based parameter ($S_{\text{AX}}$). For both parameters, the average X position resolution degrades linearly as the noise RMS increases from 1 keV to 3 keV. However, the slope of the degradation when using the integral-based parameter is approximately $43\%$ smaller than when using the amplitude-based parameter, indicating the noise robustness of the integral-based parameter.

\section{Conclusion}
This study introduces a multi-parameter-joint method for 3D position reconstruction in orthogonal-strip planar HPGe detectors. By establishing the mapping between multiple pulse shape parameters and interaction positions, this method effectively mitigates systematic biases caused by cross-dimensional interference. Simulation results demonstrate significant improvement: the maximum reconstruction bias in the Z direction is reduced from 0.4 mm to 0.02 mm in the central region and from 2 mm to 0.15 mm near the electrodes. In the X and Y directions, the maximum reconstruction bias decreases from 0.4 mm to 0.016 mm. Additionally, the use of integral-based parameters suppress the resolution degradation caused by noise, achieving resolutions of 0.07-0.16 mm (Z) and 0.07-0.44 mm (X/Y) under 1 keV RMS noise.
\par{}
The proposed multi-parameter-joint reconstruction method and the noise-resilient integral-based parameters provide an insight for  the position reconstruction of other position-sensitive detectors, such as CdZnTe detectors. It should be noted that the effectiveness of this method relies on the accuracy of pulse shape simulations. Hence, future work will focus on the experimental validation of the simulated results. The improved reconstruction performance holds significant potential for improving spatial resolution in medical imaging systems and increasing gamma-ray tracking accuracy in astronomical telescopes.

\vspace*{\fill}
\clearpage

\acknowledgments
This work was supported by the National Natural Science Foundation of China (Grants No. U1865205).


\bibliographystyle{JHEP}
\bibliography{./main}

\end{document}